\documentclass[pra,twocolumn,floatfix]{revtex4}

\usepackage{times,txfonts,graphicx}

\mathchardef\ogon="012C%
\newcommand{\as}{a\kern-0.22em\lower.40ex\hbox{$_{\ogon}$}}

\begin{document}

\title{Probing the classical field approximation -- thermodynamics and 
decaying vortices}

\author{Harry Schmidt$^{1,2}$, Krzysztof G{\'o}ral$^{1}$, Filip 
Floegel$^{1,3}$, Mariusz Gajda$^{4}$, and Kazimierz 
Rz{\as}\.zewski$^{1,5}$}
 
\affiliation{$^1$ Center for Theoretical Physics, Polish Academy of 
Sciences, Aleja Lotnik\'ow 32/46, 02-668 Warsaw, Poland}
\affiliation{$^2$ Institut f\"ur Theoretische Physik 1, Universit\"at 
Stuttgart, Pfaffenwaldring 57, D-70550 Stuttgart, Germany}
\affiliation{$^3$ Institut f\"ur Theoretische Physik, Universit\"at 
Hannover, D-30167 Hannover, Germany}
\affiliation{$^4$ Institute of Physics, Polish Academy of Sciences, Aleja 
Lotnik\'ow 32/46, 02-668 Warsaw, Poland}
\affiliation{$^5$ Cardinal Wyszy{\' n}ski University, Aleja Lotnik\'ow 
32/46, 02-668 Warsaw, Poland}


\begin{abstract}

We review our version of the classical field approximation to the
dynamics of a finite temperature Bose gas. In the case of a periodic box 
potential, we investigate the role of the high momentum cut-off, essential 
in the method. In particular, we show that the cut-off going to infinity 
limit decribes the particle number going to infinity  with the scattering 
length going to zero. In this weak interaction limit, the relative 
population of the condensate tends to unity. We also show that the 
cross-over energy, at which the probability distribution of the condensate 
occupation changes its character, grows with a growing scattering length. 
In the more physical case of the condensate in the harmonic trap we 
investigate the dissipative dynamics of a vortex. We compare the decay 
time and the velocities of the vortex with the available  analytic 
estimates.


\end{abstract}

\maketitle

\section{Introduction}

Experimental realization of Bose-Einstein condensation in a dilute gas of 
alkali atoms \cite{BEC} stimulated a number of theoretical studies of the 
quantum degenerate, weakly interacting many body system. While the lowest 
energy state of the celebrated Gross-Pitaevski equation \cite{Dalfovo} 
describes remarkably well the properties of the condensate at zero 
temperature, the dynamics of the Bose gas at non-zero temperatures remains a 
challenge. The most successful description of the temperature dependence of 
the condensate oscillation frequencies and their damping rates has been 
obtained using different versions of the two gas models \cite{Zaremba}. While 
useful, these models assume from the very beginning that the system at finite 
temperatures consists of two distinct factions: the condensate and the 
thermal cloud. A more fundamental approach would deduce the very existence of 
the condensed part directly from the dynamics of the many body system. In a 
series of papers, several groups 
\cite{semiclassical,OptExp,Burnett,Matt,PRL} proposed a classical field 
approximation to fulfill this goal. In this approximation, the bosonic 
field operator is replaced by a classical field. 
This dramatic simplification is rooted in quantum electrodynamics, where 
classical electric and magnetic fields arise naturally if the number of 
quanta is large in a 
given mode. The classical field approximation is a simple and convenient 
technique describing the condensate in dynamical equilibrium with the thermal 
cloud, at temperatures close to the critical one. We have shown  that the 
whole isolated system may be viewed as a single classical field undergoing 
nonlinear dynamics leading to a steady state \cite{OptExp,PRL} (see also 
\cite{Burnett,Matt}). The condensate is defined as the dominant term in 
the 
spectral decomposition of the time-averaged single-particle density 
matrix. Two cases were discussed: the box with periodic boundary 
conditions \cite{OptExp,Burnett,Matt} and the realistic case of a 
spherically 
symmetric harmonic trap \cite{PRL}. In our procedure it is the observation 
process and the finite detection time that allow for splitting the system 
into the condensate and the thermal cloud. The aim of the present paper is to 
further corroborate the details and the applicability of the classical fields 
approximation.

In Section \ref{method} we briefly review the approximation based on the 
classical fields. We stress again the need for the time averaging of the 
single particle density matrix for unambiguous splitting of the system into 
the condensed and uncondensed phases. The classical field approximation is 
introduced with the high momentum cut-off. In Section \ref{box} we study in 
some detail the role of this cut-off in the model of a cubic box with the 
periodic boundary conditions. We also analyze the dependence of the 
cross-over energy on the value of the coupling constant. Within our 
approximation, it is a growing function of the scattering length. In 
Section \ref{vortices}, for the first time, we apply the method to a 
dynamical dissipative process. We phase imprint the vortex on a finite 
temperature, partially condensed Bose gas and track its decay. We conclude 
with the summary in Section \ref{summary}.

\section{The method}
\label{method}

We start with a brief description of our approach (some other details can be
found also in\cite{OptExp,Burnett,Matt,PRL}). First, we consider the 
simplest 
case of a system of $N$ bosonic atoms of mass $m$ interacting via two-body 
forces. The $N$-body Hamiltonian written in the second quantization 
formalism has a form:
\begin{equation}
H=\int{\rm d}^3r \, \hat{\Psi}^{\dagger}({\bf r}){{\bf p}^2 \over 2m}
\hat{\Psi}({\bf r}) + \frac{g}{2}\int{\rm d}^3r \, 
\hat{\Psi}^{\dagger}({\bf
r})\hat{\Psi}^{\dagger} ({\bf r})\hat{\Psi}({\bf r})\hat{\Psi}({\bf r}),
\label{hamiltonian}
\end{equation}
where the first term is the kinetic energy and the second one is the 
energy of two-body interactions. In writing the Hamiltonian 
(\ref{hamiltonian}) we made a standard assumption: in the low energy limit 
a two-body interaction potential is approximated by a zero-range one. The 
interaction strength $g$ is determined by  a single parameter -- the 
s-wave scattering length $a_s$, $g=4\pi \hbar^2 a_s/m$. $\hat{\Psi}({\bf 
r})$ is the  field operator satisfying equal-time bosonic commutation 
relations:

\begin{equation}
\label{commutator}
[\hat{\Psi}({\bf r},t), \hat{\Psi}^{\dagger}({\bf r}^{\prime},t)]=\delta({\bf 
r}-{\bf r}^{\prime}).
\end{equation}
The field operator can be expanded into any complete set of
single-particle wave functions. One of these basis sets is of particular
importance -- the one which corresponds to eigenfunctions of a
single-particle density matrix\cite{definition}. This is
because these functions have direct physical interpretation as they
define coherent modes of the system. These modes are related to,
typically performed, single-particle measurements. Note that a 
theoretical description of correlated measurements, which involve a 
simultaneous detection of a few particles, should be based on an 
appropriate multi-particle reduced density matrix. The single-particle 
mode of macroscopic occupation is a Bose condensate. In the ideal gas case
eigenfunctions of the single-particle density matrix
are obviously one particle states of the external potential. The situation
becomes more complex if particles interact. In general, the eigenmodes are
unknown. The only exception is a stationary system of particles trapped in 
a box with periodic boundary conditions. Here the symmetry of the
problem imposes very strong constrains and eigenfunctions of the
single-particle density matrix are simply plane waves -- even in the
presence of interactions. The field operator is therefore:
\begin{equation}
\label{modeexpand}
\hat{\Psi}({\bf r})=\sum_{\bf k}{1 \over \sqrt{V}} e^{-i{\bf k \cdot
r}} \hat{a}_{\bf k},
\end{equation} 
where $a_{\bf k}$ is a bosonic operator which annihilates a particle of
momentum ${\bf k}$. A full operator solution of the Heisenberg equation
originating from the above Hamiltonian is not available and some 
approximations are necessary. Following Bogoliubov, at zero temperature, when 
all particles occupy a single zero-momentum mode, the corresponding 
annihilation operator can be  substituted by a c-number amplitude, 
$\hat{a}_{\bf 0} \longrightarrow \sqrt{N} a_{\bf 0}$. This is possible 
because the commutator $[\hat{a}_{\bf 0},\hat{a}_{\bf0}^{\dagger}] =1 \ll 
N$ is much smaller than the number of particles  occupying the 
zero-momentum state. The remaining terms of the expansion 
(\ref{modeexpand}) represent quantum corrections which in the lowest  
order of approximation  are neglected. Let us note that the procedure 
leads to a substitution of the the full field operator by a c-number wave 
function:

\begin{equation}
\label{c_number}
\hat{\Psi}({\bf r})\longrightarrow \sqrt{N}\Psi({\bf r}).  
\end{equation}
The condensate wave function satisfies the Gross-Pitaevskii equation
which is successfully used for description of low temperature behavior
of the Bose-Einstein condensate:
\begin{equation}
\label{GP}
i \hbar \frac{\partial \Psi({\bf r},t)}{\partial t}= 
\left({{\bf p}^2 \over 2m}+gN| \Psi({\bf r},t)|^2\right)\Psi({\bf r},t).
\end{equation}
This equation preserves both the energy and the particle number, i.e. the 
normalization of the wave function $1=\int {\rm d}^3r \, \Psi^{*}({\bf 
r})\Psi({\bf r})$. The Gross-Pitaevskii equation can also describe trapped 
condensates when a term describing the external potential is included.

Our approach is a simple generalization of the above procedure. Let us 
notice first that at relatively high energies (but below the critical one) 
there exists a number of modes, say $M$, whose occupation significantly 
exceeds unity, i.e. many different momentum states, up to a certain ${\bf 
k}_{\text{max}}$ are highly populated. Consequently, all corresponding 
annihilation operators in the expansion (\ref{modeexpand})  can be 
substituted by complex amplitudes and the remaining terms may be 
neglected:

\begin{equation}
\label{HTExp}
\hat{\Psi}({\bf r})\longrightarrow \sqrt{N}\sum_{\bf k}^{{\bf 
k}_{\text{max}}} 
\sqrt{1 \over V} e^{-i{\bf k \cdot r}} {a}_{\bf k}= \sqrt{N}\Psi({\bf r})
\end{equation}
This approach evidently leads to the same Gross-Pitaevskii equation 
(\ref{GP}) for a wave function describing the ``macroscopic part'' of the 
whole system, or, more precisely, to its finite lattice version.  This 
kind of approach is frequently used in quantum electrodynamics. The only 
difference from the zero temperature case is that now this wave function
does not correspond to the minimal energy solution. Instead, it is a 
relatively high energy state. As it has been shown 
in\cite{OptExp,Burnett,Matt}, regardless of the particular choice of an 
initial
state $\Psi({\bf r},t=0)$, different than the Gross-Pitaevskii eigenstate, 
the system evolves towards the same stationary state uniquely determined 
by the energy and the particle number. More precisely,  occupations of all 
macroscopically populated eigenmodes $N_{\bf k}=N |a_{\bf k}|^2$ fluctuate 
around  their mean  values\cite{OptExp}. The self consistency of the 
c-number approximation requires that all populations have to be large, 
$N_{\bf k}>1$. To this end, the number of classical modes $M$ has to be 
carefully chosen for each energy and particle number\cite{Burnett,Matt}. 
This 
fact can be checked only {\it a posteriori} when the stationary 
populations are obtained. The intimate link between the energy, the number 
of particles and the number of modes is the essential ingredient of the 
whole method.

There is still one point which requires additional discussion. In fact, 
the ``quasi-stationary'' high energy solution of the Gross-Pitaevskii  
equation represents a pure state of the system with {\it all particles} in 
the same single state $\Psi({\bf r},t)$. Therefore, it seems to be totally 
unjustified to interpret $|a_{\bf k}|^2$ as relative occupations 
of different coherent modes of the system. The last concept is associated 
with a mixed state. Detailed inspection shows, however, that the relative 
phases of complex amplitudes $a_{\bf k}$ vary in time rapidly. On the 
other hand an observation process lasts for a finite period of time, 
typically of the order of hundreds of microseconds. It is sufficiently 
long for a ``decoherence'' of different eigenmodes:

\begin{equation}
\label{average}
\overline{a^*_{{\bf k}_1} a_{{\bf k}_2}} \equiv {1 \over \Delta t} 
\int_{t-\Delta
t/2}^{t+\Delta t/2}{\rm d}\tau \, a^*_{{\bf k}_1}(\tau) a_{{\bf
k}_2}(\tau) \approx |a_{{\bf k}_1}|^2 \delta_{{\bf k}_1,{\bf k}_2}.
\end{equation}
In the stationary regime the time averaged quantities do not depend on
time. On the contrary, a temporal wave function of the whole system varies 
very rapidly. Therefore, what is being observed is related to a time 
averaged single-particle density matrix $\overline{\rho({\bf r}_1,{\bf 
r}_2)} = \overline{\Psi^*({\bf r}_1)\Psi({\bf r}_2)}$. It is the 
observation that reduces a pure state to a mixed one. For particles 
trapped in a periodic box we have:

\begin{equation}
\label{time_box}
\overline{\rho({\bf r}_1,{\bf r}_2)} = {N \over V} \sum_{\bf
k}^{{\bf k}_{\text{max}}}e^{i{\bf k} ({\bf r}_1 -{\bf r}_2)} |a_{\bf 
k}|^2.
\end{equation}
Eq.(\ref{time_box}) shows that eigenvalues of the time averaged
single-particle matrix are normalized occupations of different
coherent modes of the system. We want to stress that the averaging
procedure is essential for the correct physical interpretation of the high
energy solutions of the Gross-Pitaevskii equation. Note that instead of 
averaging over time, spatial avergaing can be performed. Their importance 
can be fully appreciated in a more realistic case of harmonically trapped 
Bose condensates when eigenmodes of a single-particle matrix are not 
known. It is just diagonalization of the time averaged density matrix 
which gives simultaneously the eigenmodes and the populations. Results of 
our calculations for a harmonically trapped Bose condensate at finite 
energy are presented in\cite{PRL}. The time averaged density has a 
characteristic bimodal pattern: sharp peak of condensed atoms embedded in 
a broad thermal cloud. On the other hand, an instantaneous distribution 
consists of a number of irregular  spiky structures\cite{PRL,Acta}. 
According to our numerical calculations, the average width of a single 
spike is about three times larger than the healing length, the distance 
over which the condensate wave function heals back when perturbed locally. 
The characteristic length scale of the irregular structure is slightly 
larger (but of the same order) than the grid spacing necessary for 
description of the system of a given energy and particle number. As we 
have already mentioned, the self-consistency of the approximation requires 
a very precise selection of a number of modes $M$ what in the present case 
directly translates into the grid spacing. In our approach $M$ becomes an 
important, physical parameter.

\section{Bose-Einstein condensate in a periodic box}
\label{box}

In this section we explore our approach and present some important 
technical details. We consider here the system trapped in a periodic box 
because in this situation the coherent eigenmodes are uniquely determined 
by the symmetry of the potential. This greatly simplifies the problem. We 
numerically solve the Gross-Pitaevskii equation (\ref{GP}) on a finite grid 
using the Fast Fourier Transform split-operator method. The number of grid 
points is equal to the number of different momentum states. In order to 
explore the link between the number of macroscopically occupied modes, the 
number of particles and the energy we solve Eq.(\ref{GP}) in a 3D geometry
varying the number of eigenmodes $M$ only. In the calculations we keep the
same initial state $\Psi({\bf r},t=0)$ which guarantees that the energy per 
particle is constant. Similarly, the value of the following product is fixed:

\begin{equation}
\label{limit}
gN=const
\end{equation}

The results might seem surprising: the larger the number of modes $M$ 
(with other parameters constant) the higher the occupation of the 
condensate mode $|a_0|^2 \longrightarrow 1$.

This result requires a more detailed discussion. The total energy, the number 
of modes $M$ and the product of $gN$ are the {\it only} control parameters 
in our method. The number of particles $N$ as well as the coupling 
strength $g$ do not enter the dynamical equation (\ref{GP}) separately -- 
it depends only on the product $gN$. Nevertheless, the condition of 
validity of the classical field 
approximation allows to determine the above parameters quite accurately. An 
occupation of the highest eigenmode $N_{{\bf k}_{\text{max}}}= N |a_{{\bf 
k}_{\text{max}}}|^2$ must be larger then one. We arbitrarily set this 
value to 5. 
This way, having a stationary value of a relative occupation $|a_{{\bf 
k}_{\text{max}}}|^2$, we can determine both $N$ and $g$. It occurs that 
the 
requirement of self-consistency indicates that in our calculations the number 
of particles grows with $M$ while the interaction strength $g$ decreases in 
order to satisfy the condition (\ref{limit}). Only by examining the link 
between all parameters of the classical field approximation can we fully 
interprete our data.

\begin{figure}[h]
\begin{center}
\includegraphics[width=\columnwidth,clip]{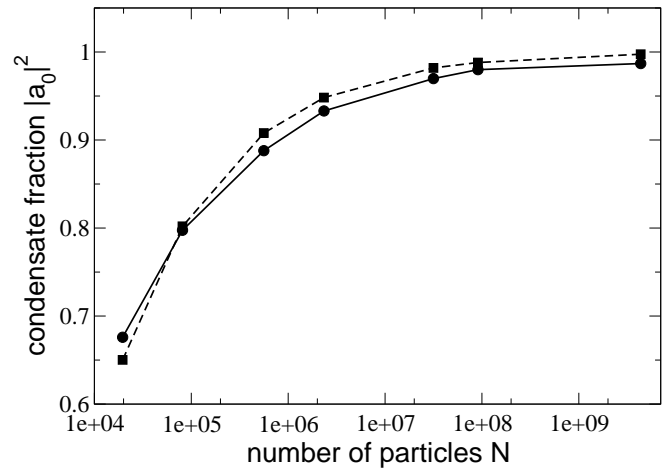}
\caption{
\label{mean_vs_Np}%
Fraction of particles in the condensate $|a_{\bf 0}|^2$ as a function of 
particle number for {\it gN=const}. For comparison we show the curve given 
by the equation $|a_0|^2 = 1-8.2/N^{1/3}$.}
\end{center}
\end{figure}

In Fig.\ref{mean_vs_Np} a relative abundance of the condensate atoms is 
plotted as a function of the number of particles. The condensate occupation 
can be quite accurately approximated by the following analytic expression:

\begin{equation}
|a_{\bf 0}|^2 \approx  1-{8.2 \over N^{1 \over 3}}.
\end{equation}

\noindent This behavior is in agreement with a recent finding\cite{Lieb}. 
Lieb and Seiringer studied the properties of a true ground state $N$-body 
wavefunction of a system of $N$ Bose particles interacting via two-body 
forces. They have shown that in the limit of $N\rightarrow \infty$ with 
$gN=const$ the relative population of the dominant eigenmode of the 
corresponding single particle density matrix aproaches $1$ while the 
eigenmode tends towards a ground state solution of the Gross-Pitaevskii 
equation. Our method shows how this limit is approached.

In the next series of calculations we estimate the interaction-induced 
shift of the cross-over energy for Bose-Einstein condensation. We prefer 
to use the term "cross-over energy" instead of "critical energy" as, 
strictly speaking, the critical energy is defined in the thermodynamic 
limit only. For a finite system (like the one we deal with here) several 
measures of the corresponding "cross-over" energy can be defined as there 
is no sharp phase transition. One of them \cite{Wilkens} is based on the 
rapid change of the probability distribution of the condensate occupation.

\begin{figure}[h]
\begin{center}
\includegraphics[width=\columnwidth,clip]{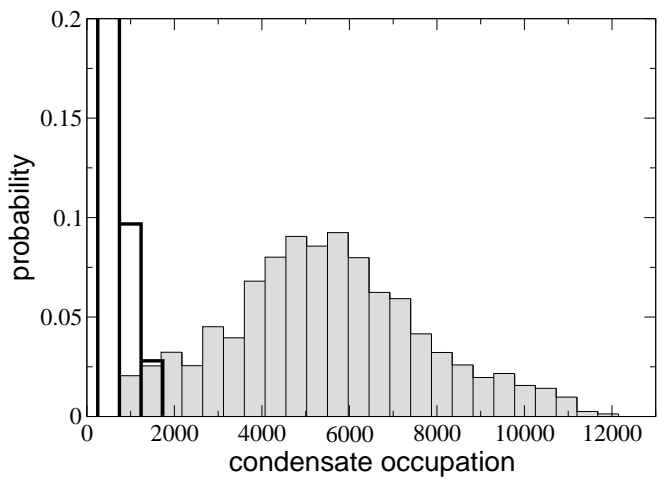}
\caption{
\label{histogram}   
Statistics of the condensate occupation in time. The grey histogram
presents
the case slightly below the cross-over energy (energy per particle
$E=43.8\hbar\omega$), while the transparent histogram corresponds to the
situation above the cross-over energy (energy per particle
$E=49.5\hbar\omega$). Note that the highest bar of the latter histogram is
off the scale of the figure (its value is 0.87). Here $N=10^5$ and
$g=10^{-4}$.}
\end{center}
\end{figure}

We study the occupation of the condensate mode $N_{\bf 0}$. The total 
energy of the system is changed while $N$ and $g$ are kept constant. This 
requires an optimization of the number of classical fields $M$ to be taken 
into account: again, we set the occupation of the least occupied mode to 
5 ($N_{{\bf k}_{\text{max}}}=5$). The above constraint does not allow for 
a continuous change of the energy though. When a quasi-stationary state of 
a given energy is reached we trace a temporal population of the condensate 
mode $N_{\bf 0}(t)$ for a long time (about $10^5$ time steps). This way we 
obtain a probability distribution of the condensate population which is 
presented in Fig.\ref{histogram}. In this figure we show two 
characteristic distributions for different energies of the system. One of 
them (for the energy per particle $E=43.8 \hbar E_1$) is centered around 
some non-zero mean value of $N_{\bf 0}\approx 5000$ (the grey histogram), 
while the other (for the energy $E=49.5 \hbar E_1$) is peaked at $N_{\bf 
0}=0$. Here $E_1=\hbar^{2}(2\pi)^2/(2mL^2)$ is the energy of the first 
excited state. 

\begin{figure}[h]
\begin{center}
\includegraphics[width=\columnwidth,clip]{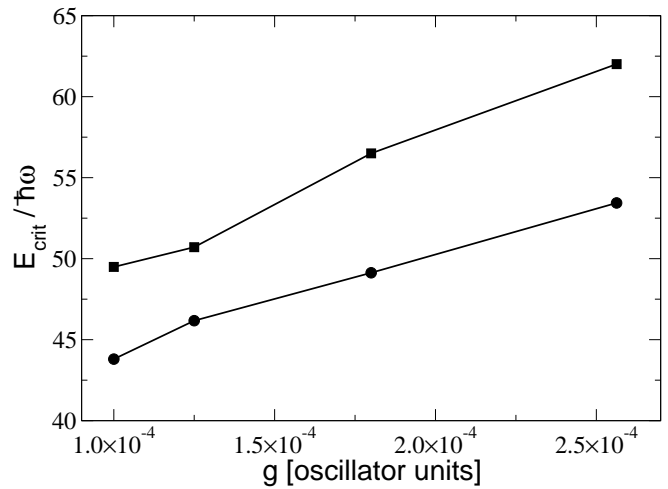}
\caption{
\label{ekrit_vs_g}
Estimates of the critical energy of the phase transition based on the
analysis of pairs of histograms like in Fig.\ref{histogram}. Total number
of
particles is $N=10^5$.
}
\end{center}
\end{figure}

Despite the fact that the values of energy for which both histograms are 
plotted are very similar, the rapid change of the character of the 
statistics evidently signifies the onset of Bose-Einstein condensation. 
The two energies give the 
lower and upper limit on the critical energy for a given interaction 
strength. We cannot determine the critical energy with better accuracy 
because of the previously discussed link between the parameters of our 
method. In the Fig.\ref{ekrit_vs_g} we show the upper and lower limits for 
the critical energy as a function of the coupling strength $g$. The critical 
energy grows with $g$. The same must happen to the critical temperature. We 
do not estimate the critical temperature as the method based on fitting the 
thermal distribution to the outer wings of the time averaged density profiles 
introduces a large uncertaintity\cite{PRL}. There is still a controversy in 
the literature\cite{Huang} about the magnitude and the sign of the shift 
of critical temperature due to interactions. The accuracy of our 
calculations, at the present stage, does not allow for a precise 
determination of this quantity. Nevertheless, they certainly indicate that 
the critical energy (and temperature) in the case of gas trapped in a box 
grows with interactions. The same conclusion for the classical field 
simulations in a box has been reached recently in \cite{Matt}.

\section{Vortex dynamics at finite temperatures}
\label{vortices}

Creation of quantum vortices has provided a long-sought clear evidence of 
supefluidity in Bose-Einstein condensates of dilute atomic gases 
\cite{JILA,Dalibard}. Following the two pioneering experiments, various 
mechanisms of vortex nucleation have been investigated \cite{mechanisms}. 
More recently, experiments have focused on peculiar properties of vortex 
lattices \cite{lattice}.

In most experiments vortices have been created in pure, essentially zero 
temperature, condensates. Very little has been said about finite temperature 
properties of these systems. Madison {\it et al.} \cite{Dalibard}, operating 
at temperatures below their detection limit, have measured finite vortex 
lifetimes of the order of 500-1000 ms. Anderson {\it et al.} 
\cite{precession} have seen no thermal damping on the time scale of 1 s. 
Finite temperature decay of vortex lattices has been also investigated 
\cite{lattice_decay}. Similarly, few theoretical works approached the 
problem. Fedichev and Shlyapnikov \cite{Gora} show that it is the 
interaction of the thermal cloud with the vortex that provides a mechanism 
of energy dissipation. Due to that, an off-center vortex spirals out to the 
condensate boundary where it decays through the creation of excitations. For 
typical experimental parameters, they estimate the vortex liftime to be in 
the range 0.1-10 s. Temperature dependence of the critical rotation 
frequency for the creation of vortices has been calculated in 
\cite{Stringari}. Bogoliubov as well as Hartree-Fock-Bogoliubov-Popov 
theories have been used to study the stability of vortex states at finite 
temperature \cite{finiteT}. The role of filling the vortex core by the 
thermal cloud as a stabilizing factor has been emphasized \cite{finiteT}.

\begin{figure}[htbp]
\begin{center}
\includegraphics[width=\columnwidth,clip]{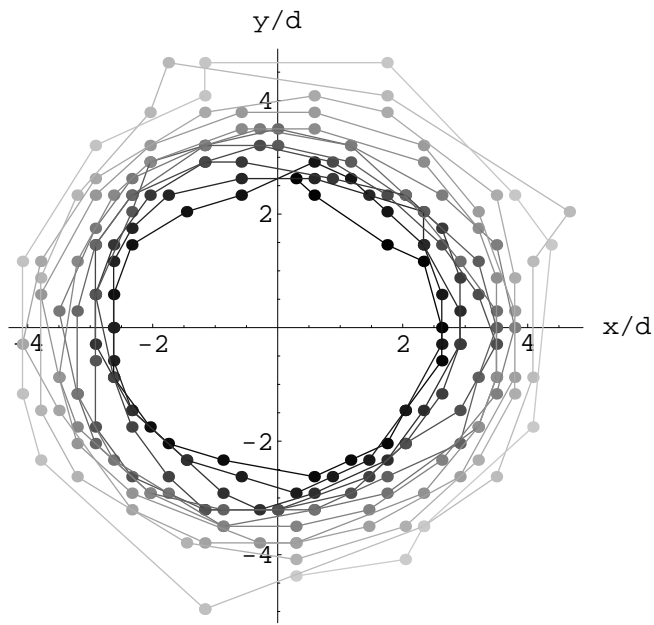}
\caption{
\label{trajectory}%
Trajectory of a vortex core. The intensity of lines and dots fades away
with time.}
\end{center}  
\end{figure}

We have used the classical field method to study the 
dynamics of vortices in the presence of a thermal cloud. As an example, we 
consider $10^5$ $^{87}$Rb atoms in a spherically symmetric trap of 
frequency $\omega=2\pi 100$ Hz. First, a suitable phase pattern 
(corresponding to a single quantum of circulation) is imprinted on a 
finite-temperature equilibrium  state of an interacting Bose gas. 
Experimentally, phase imprinting is realized by shining a far off-resonant 
light on the condensate through a glass plate whose opacity varies 
continuously with an azimuthal angle. As a result, the condensate atoms 
acquire a phase depending on their location \cite{Dobrek}. We situate 
the core of the vortex off the trap center. We see that the vortex 
created in this way starts moving around the trap center, slowly spiraling 
out to the border of 
the condensate. We follow the trajectory of the vortex core by 
time-averaging the gas density over short time. A sample trajectory of the 
vortex is shown in Fig.\ref{trajectory}. The equilibrium state used here 
for the phase imprinting contained about 48\% atoms in the condensate (its 
energy being equal to 49.46 $\hbar\omega$). The initial displacement of 
the vortex core from the trap center is $2.66 d$. The Thomas-Fermi radius 
of a condensate consisting of 48000 $^{87}$Rb atoms in our trap is $5.205 
d$ ($d=\sqrt{\hbar/m\omega}=1.09 \mu$m being an oscillator unit of length) 
-- note that the vortex disappears reaching roughly this distance from the 
trap center.

\begin{figure}[htbp]
\begin{center}
\includegraphics[width=\columnwidth,clip]{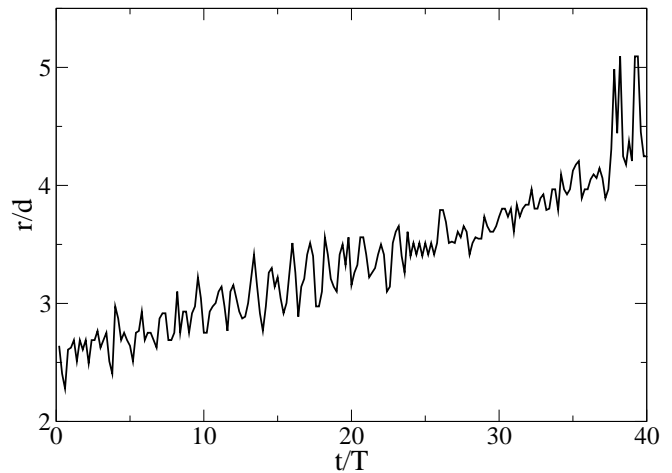}
\caption{
\label{radius}%
Time dependence of the distance of the vortex core from the trap center.
Time expressed in units of a trap period $T$.}
\end{center}
\end{figure}

Fig.\ref{radius} shows the time dependence of the distance of the
vortex core from the trap center. One can see a slow drift towards the 
condensate boundary superposed on top of small fluctuations coming from 
the interaction with the thermal part of the system. From Fig.\ref{radius} 
one can conclude that the radial velocity of the vortex is roughly 
constant. Note also the vortex lifetime $\tau$ (here equal to about $37$ 
trap periods).

\begin{figure}[h]
\begin{center}
\includegraphics[width=\columnwidth,clip]{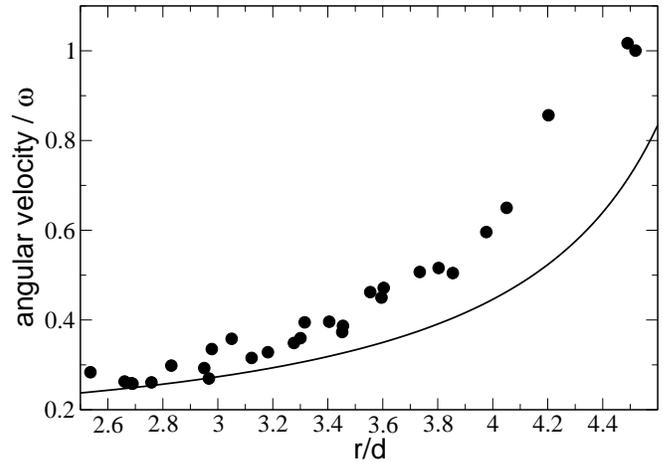}
\caption{
\label{phidot}
Angular velocity of the vortex line (expressed in units of
$\omega$) as a function of its distance from the trap center. For
comparison,
an estimate of Fetter and Svidzinsky for a pure condensate of 48000   
$^{87}$Rb
atoms in our trap is shown.}
\end{center}
\end{figure}

Fig.\ref{phidot} shows the angular velocity of the vortex core as a 
function of its (varying with time) distance from the trap center. One 
can see that the vortex moves faster and faster, in particular in the 
neighborhood of the condensate boundary. This result can be compared with 
theoretical predictions for pure, zero temperature, condensates -- 
however, in this case there is no dissipation and off center singly 
quantized vortices in our case would follow {\it closed} circular 
trajectories. For pure condensates the angular velocity of the vortex line 
diverges at the condensate boundary \cite{Fetter}. We also see this 
kind of behavior in our data for a partly condensed Bose gas. To be specific, 
we have compared our data with a variational estimate provided by Eq.(87) in 
\cite{Fetter} (here $\Omega=0$, as the trap is nonrotating) taking 48000 
$^{87}$Rb atoms. Despite notable differences between the two situations 
(presence or absence of dissipation due to the thermal cloud) the 
agreement is rather good even at the quantitative level.

\begin{figure}[htbp]
\begin{center}
\includegraphics[width=\columnwidth,clip]{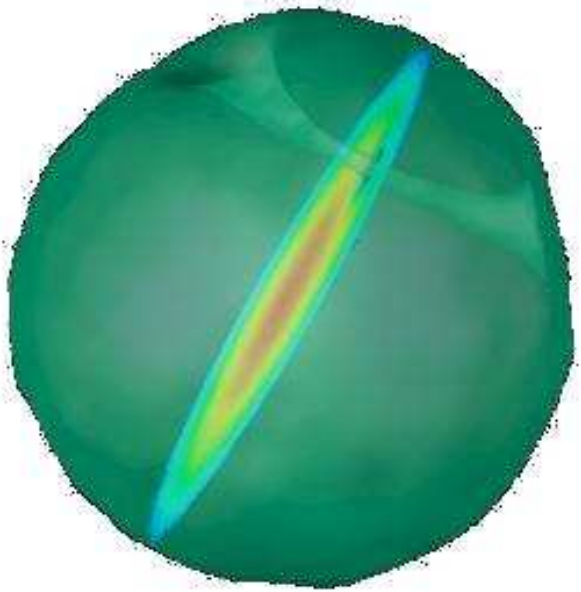}
\caption{
\label{wir}
3-dimensional density of the gas cloud with a bent vortex line. The outer
shell is a density isosurface near the condensate boundary. A plane with a
surface density plot, perpendicular to the vortex line and through the
trap center, is also shown.
}
\end{center}
\end{figure}

More features of the vortex structure can be seen in Fig.\ref{wir}. In 
this figure a 3-dimensional density of the whole gas cloud, averaged over 
a short time, is depicted. The outer shell is an isosurface of 
density near the boundary of the condensate. The whole vortex line is 
visible in the upper part of the figure: one can see that it is bent at 
the edge of the condensate and it is much thicker there than in the center 
of the trap. The former observation agrees with recent experimental and 
theoretical results \cite{bent}. A plane perpendicular to the vortex line 
and through the trap center is also shown -- it contains a surface density 
plot with a distinct hole at the vortex line.

So far we have kept the initial displacement of the vortex core fixed and 
simply studied the features of its trajectory. In the following, we 
look at the dependence of the vortex lifetime on its initial position in a 
trap. Fedichev and Shlyapnikov \cite{Gora} calculate that:

\begin{equation}
\tau=\frac{m^2 R^2 \sqrt{\mu T}}{\hbar g\rho_{n}} \ln(R/x_{min}) \, ,
\label{log}
\end{equation}

\noindent where $R$ is the spatial size of the condensate (well approximated 
by the Thomas-Fermi radius), $x_{min}$ is the initial displacement of the 
vortex line from the trap center, $T$ is temperature and $\rho_{n}\approx 
0.1 m^{5/2}T^{3/2}/\hbar^3$ is the mass density of the thermal cloud (for 
details of the derivation see \cite{Gora}; however, note the misprint in 
Eq.(12) of \cite{Gora} -- the bracketed factor should be raised to the 
power of $(-1/2)$ and not $(1/2)$ \cite{misprint}).

\begin{figure}[hbp]
\begin{center}
\includegraphics[width=\columnwidth,clip]{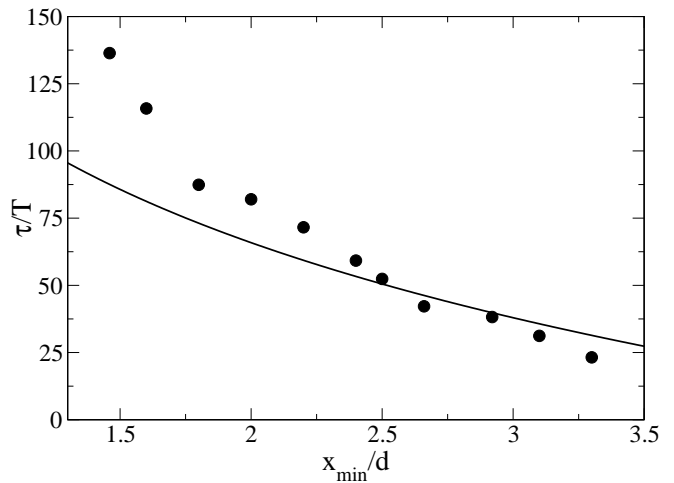}
\caption{
\label{lifetime}
Dependence of vortex lifetime $\tau$ on its initial displacement
from the  trap center $x_{min}$. Points come from our simulations while
the solid line corresponds to Eq.(\ref{log}).}
\end{center}
\end{figure}

In Fig.\ref{lifetime} we plot the vortex lifetime as a function 
of its initial position (calculated from our simulations) as well as the 
values given by Eq.(\ref{log}) (note that temperature of our system, for a 
given condensate fraction, has been estimated with the help of the 
ideal-gas formula with the finite-size corrections taken into account). 
One can see that it does agree quite well 
with the analytical estimate. However, one should note that one of the 
assumptions leading to the derivation of Eq.(\ref{log}) is that the 
condition $T \gg \mu$ holds. In our case $T/\mu=2.55$ which may be one of 
the reasons of the slight discrepances between our data and the 
analytically predicted approximate values. Note that the total energy of 
the system is still fixed to 
49.46 $\hbar\omega$.

We have also investigated the variation of the vortex lifetime with the
total energy (which amounts to its dependence on the condensate fraction).
We have seen that this relation is much weaker than the one studied above.
We have also noticed that even for a fixed initial displacement (which
does not specify its position uniquely) the lifetime changes from shot to
shot -- in other words, the complex violent behavior of the thermal cloud
in different regions of the trap makes the vortex lifetime a stochastic
parameter.

\section{Conclusions}
\label{summary}

We have reviewed our version of the classical field approximation to the
dynamics of a finite temperature Bose gas. It provides an unambiguous
splitting of the system into the condensed and uncondensed parts with the 
help of the time averaging of the single particle density matrix. In the 
case of a periodic box potential, we have investigated the role of the 
high momentum cut-off, essential in the method. In particular, we have 
shown that the cut-off going to infinity limit decribes the particle 
number going to infinity  with the scattering length going to zero. In 
this weak interaction limit, the relative population of the condensate 
tends to unity. We have also shown that the cross-over energy, at which 
the probability distribution of the condensate occupation changes its
character -- one of several measures of the energies signifying the phase
transition for the finite systems -- grows with a growing scattering 
length. In the more physical case of the condensate in the harmonic trap 
we have investigated the dissipative dynamics of a vortex. We have 
compared the decay time and the velocities of the vortex with the 
available  analytic estimates.

The next step in going beyond the approximations employed in our model is
the estimation of the influence of quantum corrections to the classical
fields, in particular the study of the corresponding time scales.

We acknowledge support from the RTN Cold Quantum Gases. M.G. acknowledges 
support by the Polish KBN grant 2 PO3B 078 19. K.R. is supported by the 
subsidy of the Foundation for Polish Science.

\end{document}